\documentclass[11pt,twoside]{article}

\usepackage{asp2006}
\usepackage{epsf}
\usepackage{psfig}
\usepackage{aas_macros}

\begin{document}
\title{Hydrogen deficient donors in low-mass X-ray binaries}   
\author{Gijs Nelemans}   
\affil{Department of Astrophysics, IMAPP, Radboud University,
  Nijmegen, the Netherlands}    

\begin{abstract} 
A number of X-ray binaries (neutron stars or black holes accreting
from a companion star) have such short orbital periods that ordinary,
hydrogen rich, stars do not fit in. Instead the mass-losing star must
be a compact, evolved star, leading to the transfer of hydrogen
deficient material to the neutron star. I discuss the current
knowledge of these objects, with focus on optical spectroscopy.
\end{abstract}



\section{Introduction: how to get hydrogen deficient donors}

In low-mass X-ray binaries (LMXBs) a compact object (either a neutron
star or a black hole) accretes material from a companion star. In the
process, copious X-rays are produced, making these objects visible to
large (many kpc) distances. Distinction based on the mass of the donor
is made between high-mass and low-mass X-ray binaries \citep[see
various chapters in][]{lk06}. Combining Kepler's law and
approximations to the maximum volume a star can occupy before starting
to transfer mass to a companion (the Roche lobe), the following
equality can be derived for stars that exactly fill their Roche lobe
\citep[e.g.][]{vh95}
\begin{equation}
P_{\rm orbital} \approx 9 \mbox{hr}
\left(\frac{\rho}{\rho_{\odot}}\right)^{-1/2}
\end{equation}
which together with the fact that the maximum density of hydrogen rich
stars (around the hydrogen burning limit) is about 100 gr/cc, implies
orbital periods for LMXBs with hydrogen rich donors larger than about 1
hour. Reversing the argument: any LMXB with a shorter orbital period
necessarily has a donor star that is more compact that a main sequence
star and thus (in view of the evolution of stellar cores) is hydrogen
deficient \citep[see][]{nrj86}. Therefore these short period LMXBs
have a special status and are called Ultra-compact X-ray binaries
(UCXBs). For a short review see \citet{nj06}. The current known sample
consists of 27 systems, 8 with known periods, 4 with tentative periods
and 15 candidate systems \citep[see][]{zjm+07}.

\section{Formation of ultra-compact (X-ray) binaries}

\begin{figure}[t]
\centerline{\psfig{figure= 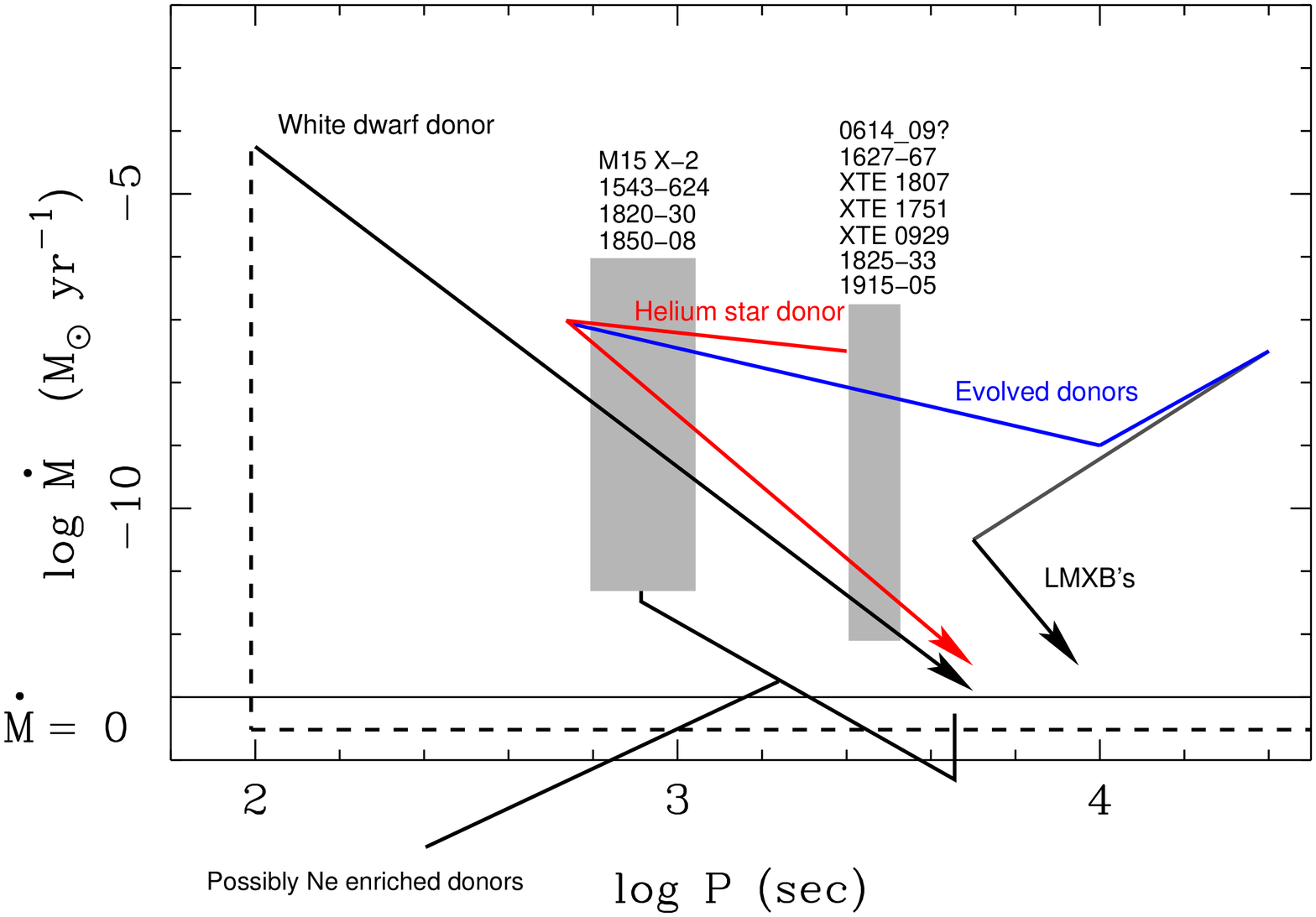,angle=-90,width=0.95\textwidth,clipping=}}
\caption{Formation paths of UCXBs stars (see text). The known systems
   are shown at their orbital period.}
\label{fig:P_Mdot}
\vspace*{-0.3cm}
\end{figure}

For a detailed discussion of the formation of UCXBs (and by analogy
their cousins with white dwarf accretors, the AM CVn systems, see
Marsh's contribution to this volume) I refer to
\citet{pw75,nrj86,skh86,tfe+87,npv+00,prp02,ynh02,svp05a}. There are
three routes for the formation of ultra-compact systems (see
Fig.~\ref{fig:P_Mdot}): (i) via a phase in which a white dwarf --
neutron stars binary or a double white dwarf loses angular momentum
due to gravitational wave radiation and evolves to shorter and shorter
periods to start mass transfer at periods of a few minutes after which
it evolves to longer periods with ever dropping mass-transfer rate
\citep[e.g.][]{ty79a}. (ii) Via a phase in which a low-mass,
non-degenerate helium star transfers matter to a neutron star or white
dwarf accretor evolving through a period minimum of about ten minutes,
when the helium star becomes semi-degenerate. After this minimum, the
periods increase again with strongly decreasing mass-transfer rate
\citep{skh86}.  (iii) From LMXBs (or cataclysmic variables) with
evolved secondaries \citep{tfe+87}, which, after mass loss of the
evolved star has uncovered the He-rich core, evolve rather similar to
the helium star tracks.  These formation paths are depicted in
Fig.~\ref{fig:P_Mdot}, together with the observed UCXBs at their
orbital periods. It is clear that in order to distinguish between the
different evolutionary scenarios, more information than the orbital
period is needed and that for the currently observed systems all
formation scenarios in principle are viable.

\section{Chemical composition of the donor}

In order to distinguish the different formation channels, the chemical
composition of the donor star is needed. For the three channels
described above the expected chemical compositions are:
\begin{description}
\item[White dwarf channel] depending on the nature of the white dwarf
  the transferred material will be mainly He with CNO processed
  (i.e. mainly N) material for a He-core white dwarf, or a C/O mixture
  in case the donor is a C/O-core white dwarf.
\item[Helium star channel] He with little N, plus possibly helium
  burning products (i.e. C and O) depending on the amount of helium
  burning that has taken place before the donors fills its Roche lobe.
\item[Evolved secondaries] He plus CNO processed material and,
  depending on the exact evolutionary history and the phase of the
  evolution, some H.
\end{description}

\begin{figure}[t]
\centerline{\psfig{figure=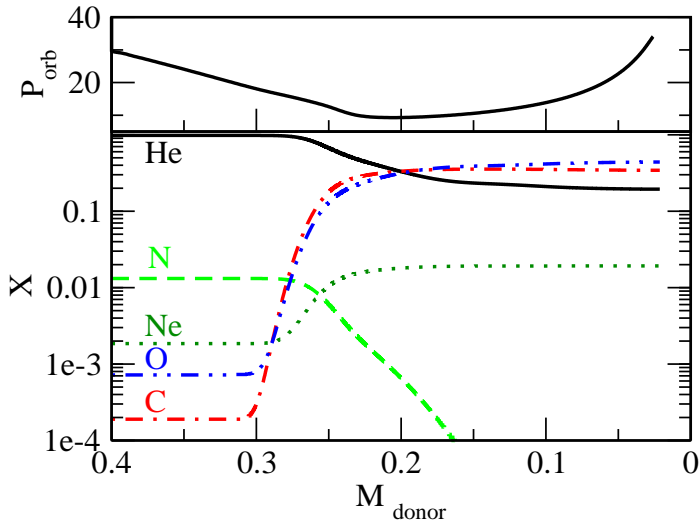,angle=-90,width=0.7\textwidth,clipping=}}
\vspace*{-0.5cm}
\caption{Orbital period (top) and chemical composition (bottom) as
  function of donor mass (and thus time, running from left to right)
  for a helium star donor that has burnt most of its helium before
  mass transfer starts. Solid line: He, dashed: N, dash-dot: C,
  dash-dot-dot: O, dotted: Ne.}
\label{fig:M_P_X}
\end{figure}

In order to asses the likelihood of helium burning products to show up
in the transferred material in helium star donors, Lev Yungelson and I
started a study of these stars. In Fig.~\ref{fig:M_P_X} we show one
example, for a case where the helium star has burnt most of its helium
before the mass transfer starts. Rapidly the helium burning products
(C, O Ne) become visible. Details will be presented elsewhere
(Nelemans \& Yungelson, in prep.) The question now is, whether
observations can tell us the chemical composition of the donors.

\section{Observational results (optical/X-ray)}

\begin{figure}[t]
\centerline{\psfig{figure=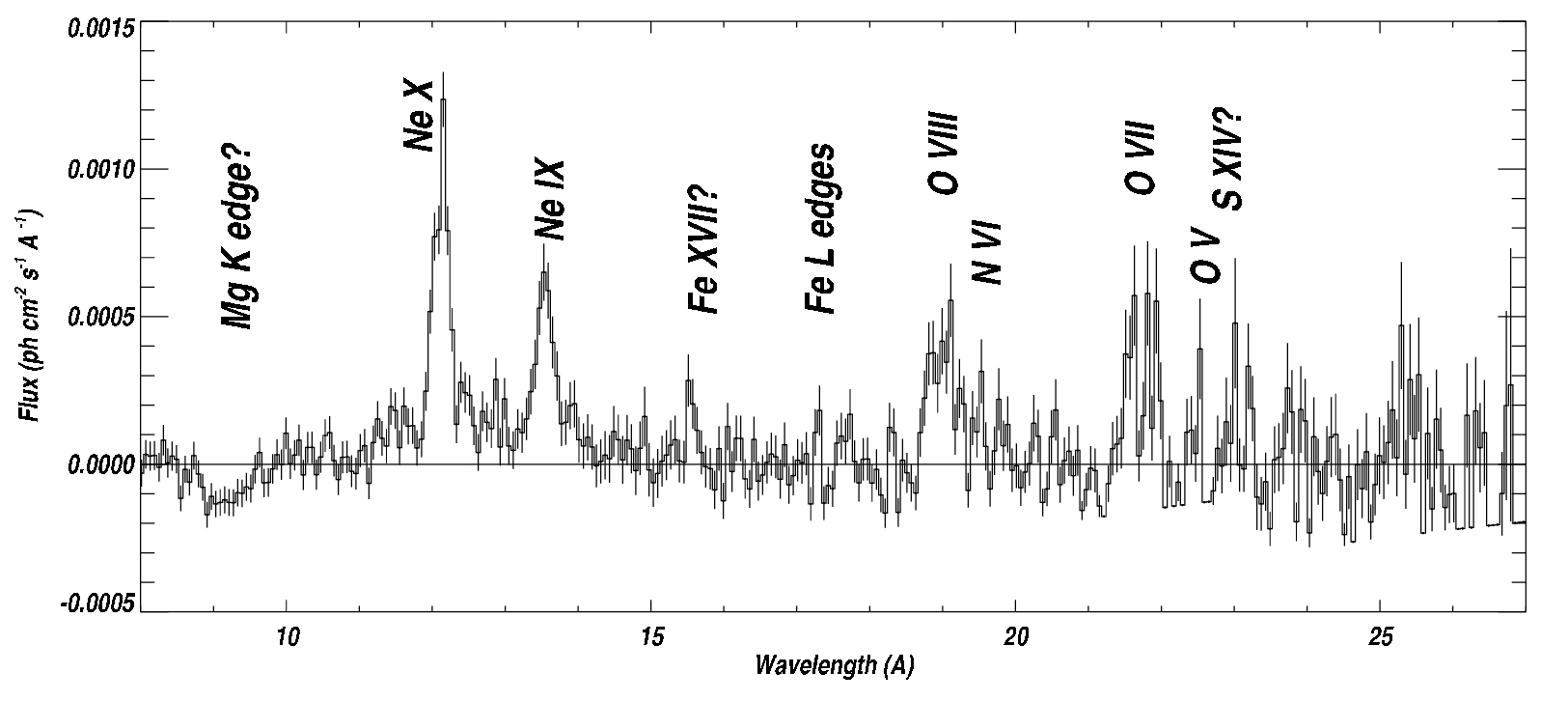,angle=0,width=0.8\textwidth,clipping=}}
\caption{X-ray spectrum of 4U~1626-67 showing double peaked lines of O
and Ne. From \citet{sch+01}}
\label{fig:1626_X}
\end{figure}

\textbf{X-ray spectra}\\ The X-ray spectra lacked resolution until
recent observations obtained with the gratings on board the {\it
Chandra} satellite.  The X--ray spectra are difficult to interpret due
to the lack of clear emission features (except in the 7s accreting
pulsar 4U 1626-67 which shows strong O and Ne emission lines in its
X-ray spectrum, \citet{sch+01} see Fig.~\ref{fig:1626_X}) and
contributions of interstellar absorption. A number of UCXBs have been
(first) identified based on their X-ray spectra, when \citet{jpc00}
noted a similarity between the spectrum of 4U 1850-087 and three more
LMXBs.  The common feature in the X-ray spectra was suggested to be
due to an enhancement of neon in these systems, however, this is
uncertain \citep[e.g.][]{jc03,wnr+06}. These observations were
interpreted as evidence for the donors being ONe, or more likely CO
white dwarfs in which Ne has sunk to the core in an earlier
evolutionary phase and is now exposed, after accretion has peeled--off
the outer layers \citep{ynh02}. Recently it has been noted that if it
is really the O/Ne ratio that is anomalous, and not the Ne abundance
itself, this points towards a He white dwarf donor, in which the O
abundance is strongly reduced due to CNO processing
\citep{icv05}. Interestingly, one system that shows a similar feature
in its X-ray spectrum \citep[4U~1556-605][]{ffm+03}, shows strong
hydrogen and helium emission in its optical spectrum, suggesting that
it is not an UCXB \citep{njs06}.

\vspace*{0.5cm}
\noindent \textbf{Type I X-ray bursts}\\ For a number of UCXBs type I X-ray
bursts have been found, especially for the ones in globular clusters,
but that may be a selection effect. The peculiar chemical composition
of the accreting material will influence the burst properties and in
principle can give an extra tool to study the chemical composition
\citep[e.g.][]{cb01}. Recently a number of peculiar bursts have been
observed in 2S~0918-549 and 4U~0614+09 \citep{icv05,kuu05} which
\citet{icv05} suggest are due to thick layers of He that are
burned. It is clear that there is still uncertainty how we should
interpret the different pieces of information, because in some cases
the bursts and the optical spectra suggest conflicting compositions as
we will discuss below!

\begin{figure}[t]
\centerline{\psfig{figure=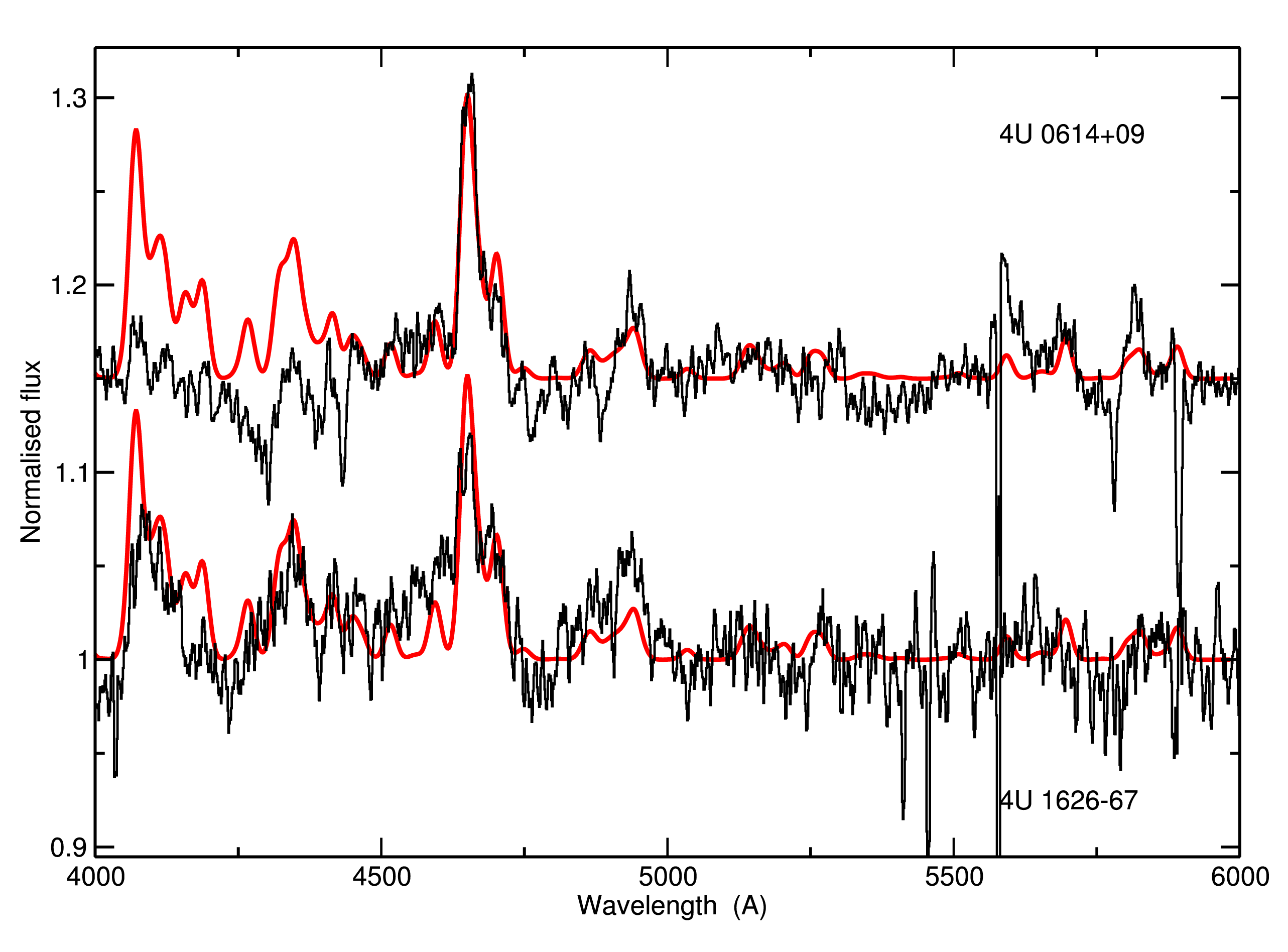,angle=0,width=0.7\textwidth,clipping=}}
\centerline{\psfig{figure=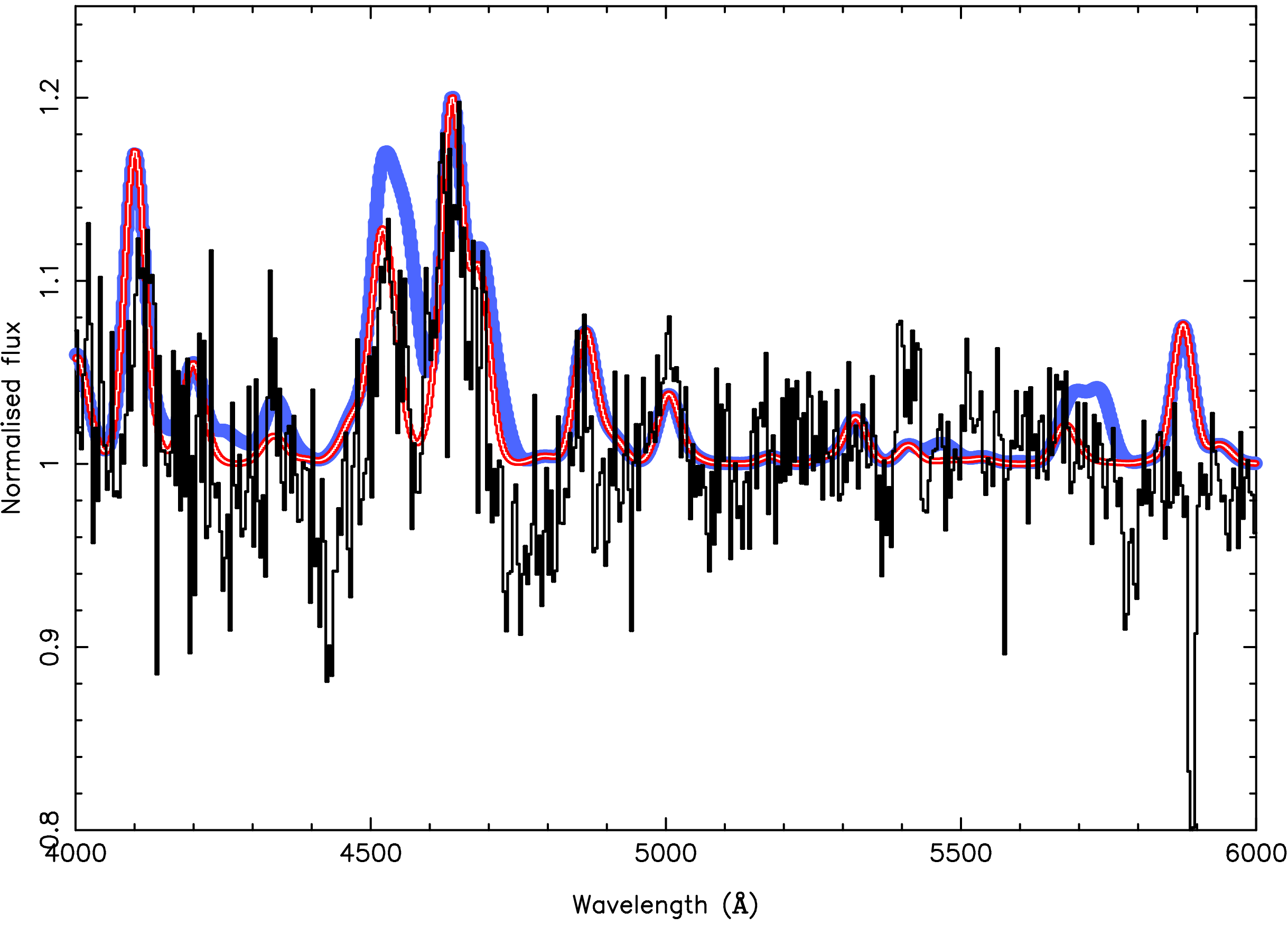,angle=-90,width=0.7\textwidth,clipping=}}
\caption{Optical spectra of 4U~0614+09 and 4U~1626-67 (top, with LTE C/O
  disc models) and 4U~1916-05 (bottom, with LTE He + N disc
  model). See \citet{njs06} for more details.}
\label{fig:UCXBs}
\vspace*{-0.3cm}
\end{figure}

\vspace*{0.5cm}
\noindent \textbf{Spectra}\\ With the advent of 8m class telescopes it has
become possible to obtain optical spectra of some (candidate) UCXBs.
We started a systematic spectroscopic study of (candidate) UCXBs using
the VLT. The main aim was to confirm/reject candidates and to study
the chemical composition of the donors stars in these systems. The
first results are published in \citet{njm+04} and can be summarised as
follows. We identified the features in the spectrum of 4U 0614+09 as
relatively low ionisation states of carbon and oxygen. This clearly
identifies this system as an UCXB and suggests the donor in this
system is a carbon-oxygen white dwarf. The similarity of the spectrum
of 4U 1543-624 suggests it is a similar system, while for 2S 0918-549
the spectrum didn't have a high enough S/N ratio to draw firm
conclusions, but it is also is consistent with being a similar system
(and clearly does not show the characteristic strong hydrogen emission
lines of low-mass X-ray binaries). We therefore concluded that all
these systems are UCXBs. The second set of observations is presented
in \citet{nj06} and \citet{njs06} and is briefly discussed here.

\begin{figure}[t]
\centerline{\psfig{figure=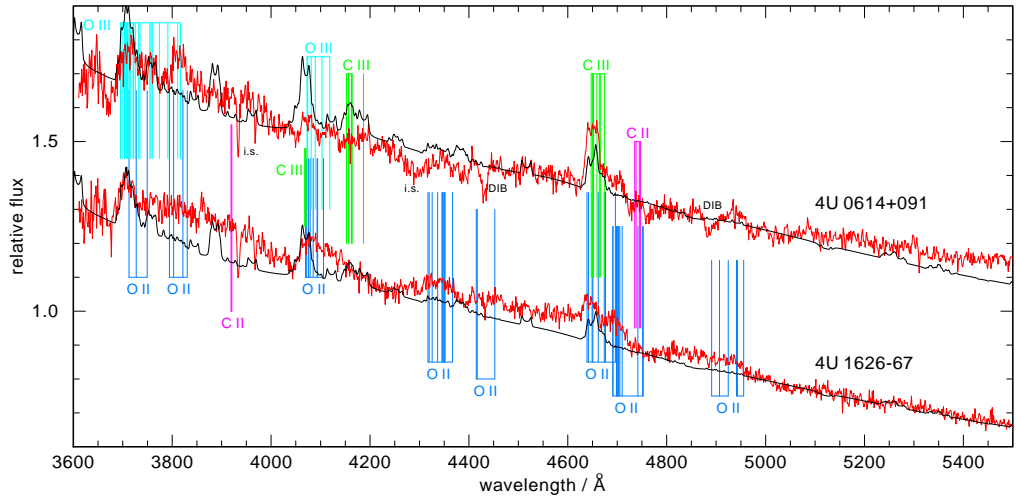,angle=0,width=1.0\textwidth,clipping=}}
\caption{Optical spectra and detailed NLTE models for 4U~0614+09 and
  4U~1626-67 from \citet{wnr+06}. }
\label{fig:Werner}
\end{figure}

For two objects (4U~0614+09 and 4U~1626-67) the are clear indications
that the discs are dominated by C and O (Figs.~\ref{fig:UCXBs} (top)
and \ref{fig:Werner}). \citet{wnr+06} have also obtained VLT spectra
of these sources and have compared these with detailed NLTE models for
the spectra of UCXBs (Fig.~\ref{fig:Werner}). Unfortunately the NLTE
models do not agree with the observed spectra in enough detail to be
able to get quantitative abundances from this analysis. Surprisingly
simple LTE models seem to fit better (Fig.~\ref{fig:UCXBs}), but can
also not be used for quantitative measurements, as NLTE effects,
mainly due to X-ray irradiation have to be important. Another surprise
is that the interpretation of the donors being C and O rich is
difficult to reconcile with the presence of type I X-ray bursts, as
they seem to require H or He to be present \citep[see][and references
therein]{icv05}. Either spallation of heavier elements
\citep[e.g.][]{bsw92} takes place, or there is He in addition to C and
O which implies a helium star donor (see Fig.~\ref{fig:M_P_X}), or the
X-ray bursts for some reason can be powered by C or O.

Another surprise came with the spectrum of 4U~1916-05, which shows
similar weak lines as the other UCXBs (Fig.~\ref{fig:UCXBs} bottom),
but can best be fitted by a model consisting of He and overabundant N
(again, because of the LTE nature of the model, no quantitative
results can be obtained). Contrary to the He dominated spectra in AM
CVn systems (see Marsh, this volume) that show very strong emission
lines, the weakness of the lines makes distinction between He and C/O
rich systems a matter of rather high signal-to-noise measurements.

\begin{figure}[t]
\centerline{\psfig{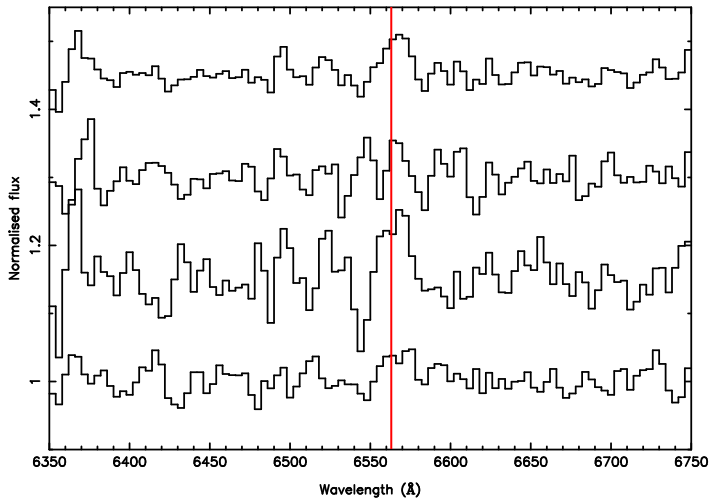}}
\caption{Optical spectra of XTE J0929-314, where the presence of
  H$\alpha$ is claimed by \citet{ccg+02}. See \citet{njs06} for a
  detailed discussion.}
\label{fig:0929}
\end{figure}

The only object where possibly traces of hydrogen are detected is the
transient XTE~J0929-314 for which \citet{ccg+02} claimed detection of
H$\alpha$ in spectra taken during the outburst. However,
Fig.~\ref{fig:0929} shows these spectra (taken from the ESO archive)
and the detection is at best marginal. This conclusion is strengthened
by the fact that in the same spectra quite clear emission around
4600\AA, which led us to tentatively conclude it is a C/O rich system
\citep{njs06}.

\section{Conclusions}

Ultra-compact X-ray binaries are puzzling end products of binary
evolution and despite substantial progress in recent years both in
observational as well as theoretical work it is not yet clear how the
observed systems are formed. The (detailed) chemical composition of
the transferred material will be a key observable to distinguish
different formation channels. In principle optical and X-ray
spectroscopy in combination with detailed modelling of the donor stars
and the disc spectra can provide this information. However of the
currently 27 (candidate) systems that are know, for only 4 or 5 we
know their (rough) chemical composition: 3 (or 4) C/O discs and 1 (or 2)
He discs.

\acknowledgements 
I wish to thank Lev Yungelson, Peter Jonker and Danny
Steeghs for a lot of joint projects that form the basis of this
article.




\end{document}